\documentclass[12pt, a4paper]{article}
\pdfoutput=1

\usepackage{amssymb}
\usepackage{amsmath}
\usepackage{amsfonts}
\usepackage{graphicx}
\usepackage{mathrsfs}
\usepackage{comment}
\usepackage{cite}
\def\->{\rightarrow}
\def\<-{\leftarrow}

\newcommand{\Slash}[1]{{\ooalign{\hfil#1\hfil\crcr\raise.167ex\hbox{/}}}}

\newcommand{\beq}{\begin{equation}}  \newcommand{\eeq}{\end{equation}}
\newcommand{\bef}{\begin{figure}}  \newcommand{\eef}{\end{figure}}
\newcommand{\bec}{\begin{center}}  \newcommand{\eec}{\end{center}}
  
\newcommand{\laq}[1]{\label{eq:#1}}  

\newcommand{\Eq}[1]{Eq.\,(\ref{eq:#1})}
\newcommand{\Eqs}[1]{Eqs.\,(\ref{eq:#1})}
\newcommand{\eq}[1]{(\ref{eq:#1})}

\newcommand{\vev}[1]{ \left\langle {#1} \right\rangle }

\newcommand{\lac}[1]{\label{chap:#1}}
\newcommand{\SU}[1]{{\rm SU{#1} } }

\newcommand{\bed}{\begin{description} \item}
\newcommand{\eed}{\end{description}}

\def\({\left(}
\def\){\right)}

\def\O{\mathcal{O}}
\def\U{\mathop{\rm U}}

\def\a{\alpha}
\def\b{\beta}

\def\d{\delta}

\def\f{\phi}
\def\g{\gamma}
\def\h{\theta}
\def\k{\kappa}
\def\l{\lambda}

\def\p{\psi}

\def\r{\rho}

\def\G{\Gamma}

\def\L{\Lambda}

\def\P{\Psi}

\def\*{\dagger}

\newcommand{\AND}{~{\rm and}~}
\newcommand{\EV}{ {\rm\, eV} }

\newcommand{\MEV}{ {\rm\, MeV} }
\newcommand{\GEV}{ {\rm\, GeV} }
\newcommand{\TEV}{ {\rm\, TeV} }

\renewcommand{\thefootnote}{\fnsymbol{footnote}}

\begin{document}
\begin{titlepage}
\begin{center}

\hfill TU-1082,~IPMU19-0026\\

\vskip .75in

{\Large\bf 
ALP inflation and Big Bang on Earth
}

\vskip .75in

{ \large   Fuminobu  Takahashi\,$^{a,b}$\footnote{email: fumi@tohoku.ac.jp},
    Wen Yin\,$^{c}$\footnote{email: yinwen@kaist.ac.kr}}

\vskip 0.25in

\begin{tabular}{ll}
$^{a}$ &\!\! {\em Department of Physics, Tohoku University, }\\
& {\em Sendai, Miyagi 980-8578, Japan}\\[.3em]
$^{b}$ &\!\! {\em Kavli IPMU (WPI), UTIAS,}\\
&{\em The University of Tokyo,  Kashiwa, Chiba 277-8583, Japan}\\[.3em]
$^{c}$ &\!\! {\em Department of Physics, KAIST, Daejeon 34141, Korea}\\[.3em]
& {\em }

\end{tabular}


\begin{abstract}
We study a hilltop inflation model where an axion-like particle (ALP) plays the role of the inflaton.
We find that, for a broad class of potentials,
the decay constant and the mass at the potential minimum satisfy the relation, $m_\phi \sim 10^{-6} f$, 
to explain the CMB normalization. The ALP is necessarily coupled to the standard model particles for successful reheating. The ALP with the above relation can be searched at beam dump experiments, e.g., the SHiP experiment, if the inflation scale is sufficiently low. In this case, the ALP decays through the interactions that led to the reheating of the Universe. 
In other words, the Big Bang may be probed at ground-based experiments. 
\end{abstract}

\end{center}
\end{titlepage}
\setcounter{footnote}{0}
\setcounter{page}{1}

\renewcommand{\thefootnote}{\arabic{footnote}}

\section{Introduction}

There is accumulating evidence for inflation in the early Universe~\cite{Starobinsky:1980te,Guth:1980zm,Sato:1980yn,Linde:1981mu,Albrecht:1982wi}. Not only does the inflation solve various initial condition problems of the standard big bang cosmology but also explains the origin of density perturbations. The observed CMB temperature anisotropy retains coherence that extends beyond the horizon at the recombination, which strongly indicates that our Universe experienced an accelerated expansion in the past.

 In the slow-roll inflationary paradigm, the flatness of the potential is essential for both driving the inflationary expansion and generating density perturbations. In particular, if the inflation scale is lower than the GUT scale, the inflaton potential must be extremely flat, which may call for some explanation. Such flat potential can be ensured by shift symmetry, in which case one can identify the inflaton with an axion or an axion-like particle (ALP).
 
Let us suppose that the inflaton, $\phi$, respects discrete shift symmetry, 
\begin{align}
 \label{pq}
    \phi \to \phi + 2\pi f,
\end{align}
where $f$ is the decay constant. Then,
the inflaton potential is periodic with periodicity $2 \pi f$:
\begin{align}
    V(\phi) = V(\phi + 2 \pi f).
\end{align}
Such periodic potential can be expanded in Fourier series
as a sum of cosine functions. 
If one of the cosine terms dominates the potential, the inflation model is reduced to the so-called 
natural inflation~\cite{Freese:1990rb,Adams:1992bn}.
However, this simple possibility is on the verge of exclusion according to the latest CMB data~\cite{Akrami:2018odb}.
Moreover, the model requires a super-Planckian decay constant, which may not be justified in a context of quantum gravity.
Therefore, some extension of the natural inflation is necessary.

Let us consider the so-called multi-natural 
inflation~\cite{Czerny:2014wza, Czerny:2014xja,Czerny:2014qqa,Higaki:2014sja}, 
where multiple cosine terms conspire to make the potential flat enough for the slow-roll inflation to take place.
In particular, we will focus on the minimal case in which the potential consists of the two cosine terms,
\begin{align}
\label{eq:DIV} 
V(\phi) = \Lambda^4\(\cos\(\frac{\phi}{f} + \theta \)- \frac{\kappa }{n^2}\cos\(\frac{n\f }{f }\)\)+{\rm 
const.}.
\end{align}
where $n$ ($>1$) is an integer,  $\kappa$ and $\theta$ parameterize the relative height and phase of the two terms, 
respectively, and the last constant term is introduced to make the cosmological constant vanishingly small in the 
present vacuum.\footnote{\label{ft1} One can consider a case in which the first cosine term also contains
another integer $m (< n)$. It is straightforward to extend our analysis to this case by redefining the decay constant. Effectively, the coefficient $n$ should be replaced with a common fraction $n/m$.
Our main results do not change significantly 
unless $m$ is significantly larger than unity.} 
In the limit of $\theta=0$ and $\kappa = 1$, the above potential is reduced to the hilltop quartic inflation model. The observed scalar spectral index is correctly reproduced by slightly varying both parameters around their canonical values \cite{Takahashi:2013cxa, Czerny:2014wza, Daido:2017wwb, Daido:2017tbr}. The flat top potential with multiple cosine terms like (\ref{eq:DIV}) has 
several possible UV completions e.g. in 
supergravity\cite{Czerny:2014xja,Czerny:2014qqa,Higaki:2014sja} and
extra dimensions~\cite{Croon:2014dma}. A similar potential with an elliptic function
is also obtained in the low-energy limit of some string-inspired set-up~\cite{Higaki:2015kta, Higaki:2016ydn}.

The inflation model with the potential (\ref{eq:DIV}) can be broadly classified into two categories depending on the parity of $n$.
While the inflaton dynamics during  inflation is not so sensitive to the value of $n$, the evolution of the Universe
after inflation crucially depends on its parity.
If $n$ is an odd integer, the inflaton potential (\ref{eq:DIV}) has an upside-down symmetry, and the inflaton
potential is approximated well by the quartic potential soon after inflation. As a result, the inflaton mass at 
the potential minimum is much smaller than the typical curvature scale of $\sim \Lambda^2/f$, and 
is actually of order the Hubble parameter during inflation. 
Such a light inflaton raises an interesting possibility that a single ALP explains both inflaton and dark matter simultaneously,
which is dubbed the ALP miracle~\cite{Daido:2017wwb, Daido:2017tbr}. Daido and the present authors
studied the detailed reheating process and found a viable parameter space which is within the sensitivity reach 
of the IAXO experiment~\cite{Irastorza:2011gs, Armengaud:2014gea, Armengaud:2019uso}, TASTE~\cite{Anastassopoulos:2017kag}, and laser-based photon colliders~ \cite{Hasebe:2015jxa,Fujii:2010is,Homma:2017cpa}.
On the other hand, if $n$ is an even integer (or a common fractional, cf. the footnote~\ref{ft1}), the inflaton is massive at the potential minimum and its mass is of order $\Lambda^2/f$.
This case was studied in e.g. Refs.~\cite{Czerny:2014wza,Czerny:2014xja,Czerny:2014qqa,Higaki:2014sja}
with a main focus on the parameter region where inflation scale is so high that observable primordial tensor modes are generated.
In this paper, we focus on its low-scale limit and explore a possibility that the inflaton or ALP may be probed 
by ground-based experiments. As we shall see later, the results obtained for even $n$ hold for a broader class of multi-natural inflation with more than two cosine terms.

In this paper we study the ALP inflation with a hilltop potential where the potential consists of multiple cosine terms. First we derive a relation between the ALP mass and decay constant, $m_\phi \sim 10^{-6} f$, from the CMB normalization condition. This relation is rather robust and it holds for a broad class of the multi-natural inflation. Indeed, essentially the same relation was found by Czerny, Higaki and one of the present authors (FT) in Ref.~\cite{Czerny:2014xja}, but its experimental 
implication was not discussed as they mainly studied high-scale inflation.
Next we explore a possibility that the ALP is within the reach of ground-based experiments such as the SHiP experiment \cite{Anelli:2015pba,
Alekhin:2015byh,
Dobrich:2015jyk, Dobrich:2019dxc}.
To this end, we study the reheating process of the inflaton for different choices of the couplings to the standard model (SM) particles, and discuss if the relation between the ALP mass and decay constant can be experimentally confirmed. Lastly we will study an inflection point inflation and show that it does not lead to successful reheating for the inflaton parameters within the experimental reach. We will also see that the inflection point inflation cannot start from the eternal inflation in contrast to the hilltop inflation. We also discuss a possible UV completion in which 
the decay constant is related to the soft supersymmetry (SUSY)
breaking.

The rest of this paper is organized as follows. In Sec.~\ref{sec:2} we revisit the multi-natural inflation with two cosine terms and derive relations between the ALP mass and decay constant. In Sec.~\ref{sec:3} we discuss the inflaton hunt and the reheating process for two different choices of the ALP couplings to the SM particles. In Sec.~\ref{sec:4} we study the inflection point inflation and its implications. We briefly discuss whether eternal inflation can be embedded in the framework in Sec.~\ref{sec:5}.
The last section is devoted to discussion and conclusions.

\section{Axion hilltop inflation}
\label{sec:2}

\begin{figure}[!t]
  \begin{center}
   \includegraphics[width=75mm]{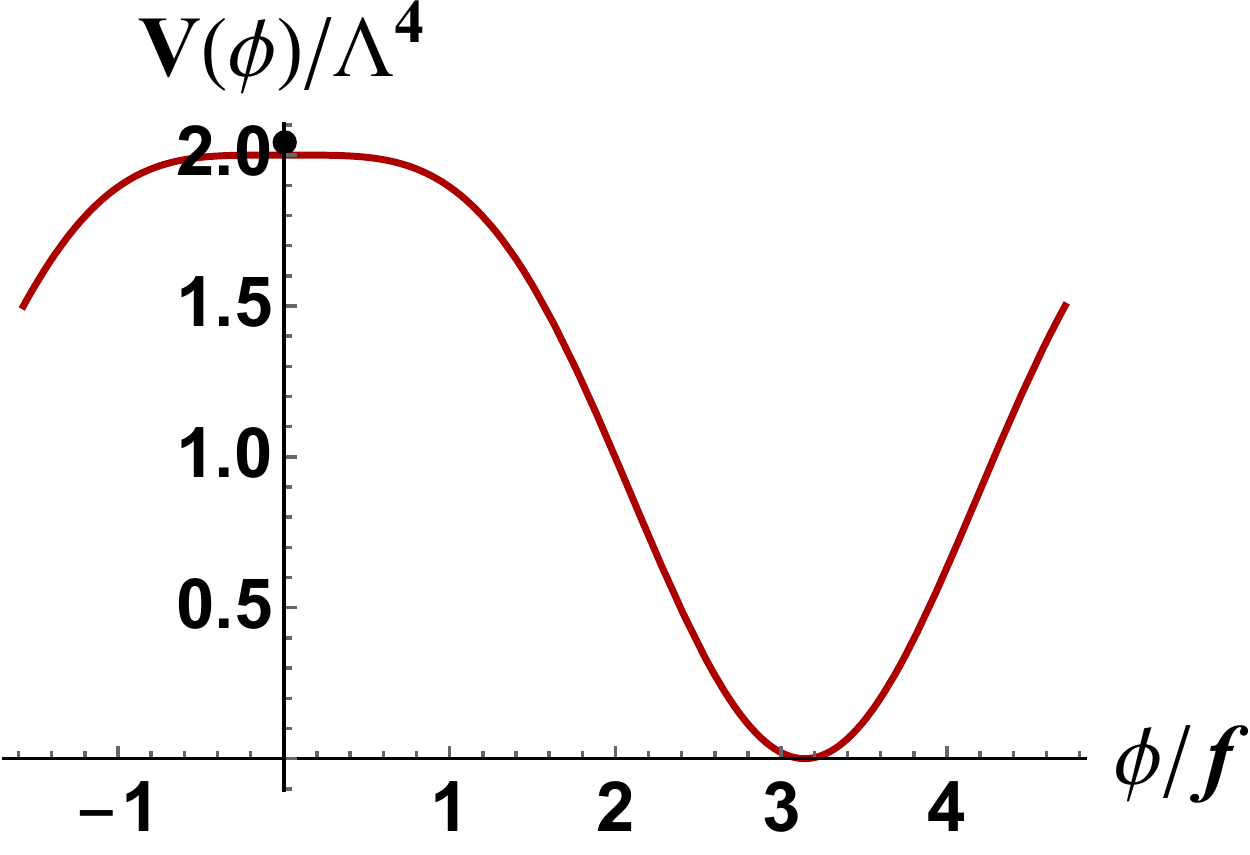}
      \includegraphics[width=75mm]{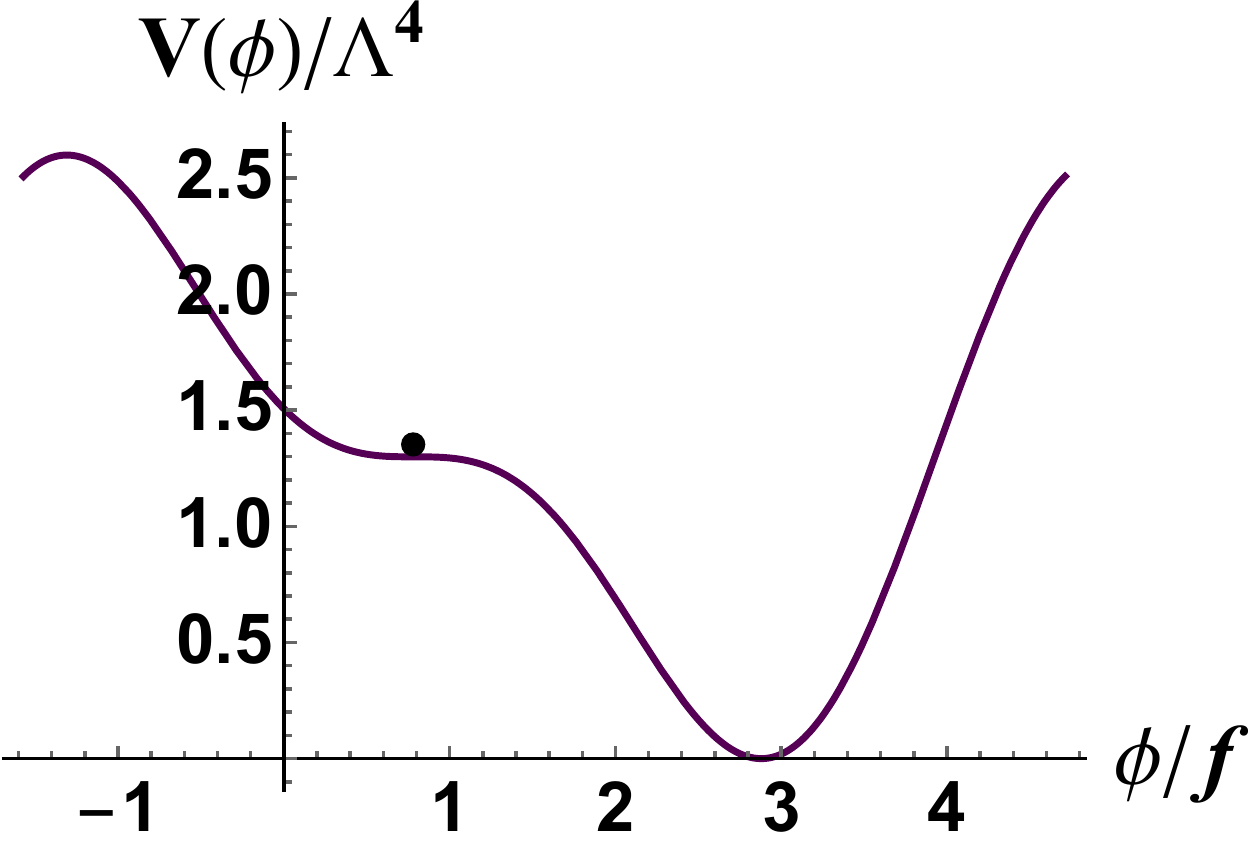}
\end{center}
\caption{
The inflaton potential of the hilltop inflation [left] and the inflection point inflation [right]. 
The slow-roll inflation is possible around the black dots. We take $(n,\theta,\kappa) = (2,0,1)$ and $(2,\pi/4,2)$ for the left and right panels, respectively.}
\label{pot}
\end{figure}

Now let us study the inflation with the potential (\ref{eq:DIV}).
As mentioned before, the inflaion is reduced to the hilltop quartic inflation model
for $\theta \simeq 0$ and $\kappa \simeq 1$. See the left panel of Fig.~\ref{pot} for the shape of the hilltop potential. In this section we focus on this case and 
show that the $\L$ and $f$ satisfy a certain relation due to the CMB normalization condition. 
Although it is possible to realize an inflection point inflation with $\theta=\O(1)$, 
 the inflation parameters for this case turn out to be well beyond the experimental reach, as we shall 
 see later in this paper.

We suppose that  the inflaton initially stays in the vicinity of the potential maximum,
which is located near the origin for $\theta \simeq 0$.
Around the origin the potential can be expanded as
\begin{equation}
\label{eq:app}
V(\phi) = V_0
- \theta \frac{ \L^4}{f} {\phi}+  \frac{(\kappa-1)}{2}\frac{\L^4}{f^2} \phi^2
+ \frac{\h }{3!}\frac{\Lambda^4 }{f^3} {\phi^3} -  \frac{n^2-1 }{4!}\(\frac{\Lambda}{f}\)^4 \phi^4  + \cdots,
\end{equation}
where $V_0$ is a constant defined below, and we have included only terms up to the first order in
$\h $ and $\kappa-1$, assuming they are much smaller 
than unity. In fact, the cubic term has a negligible effect on the inflaton dynamics during the hilltop inflation.
Then, the potential can be well approximated by
\begin{align}
\label{app}
V(\phi) \simeq V_0   - \h \frac{\L^4 }{ f} {\phi} +\frac{m^2 }{2}\f^2 - \lambda \phi^4,    
\end{align}
where we have defined 
 \begin{align}
 \label{eq:V0}
 V_0 &= \beta_n \Lambda^4 
 \equiv \left(2 - \frac{2}{n^2} \sin^2{\frac{n \pi}{2}} \right) \Lambda^4,\\
 m^2 &\equiv {\(\kappa-1\)}\frac{\L^4 }{f^2},\\
 \lambda &\equiv \frac{n^2-1 }{4!}\(\frac{\Lambda}{f}\)^4.
 \label{eq:lambda}
\end{align}
Here $\beta_n$ is a constant of order unity and defined in such a way that the potential vanishes at the minimum,
$\phi_{\rm min} \simeq \pi f$.
Obviously, the potential (\ref{app}) is reduced to that of the hilltop quartic inflation
in the limit of $\h \rightarrow 0$ and $\kappa \rightarrow 1$.
In the following analytic estimate we drop $\k-1$ and $\h$ for simplicity but they are taken into consideration
in the numerical calculations.

Let us solve the inflaton dynamics during inflation. The inflation takes place when the inflaton sits near the origin
where the potential is extremely flat. We assume $\dot{\phi} > 0$ during inflation without loss of generality. 
The flatness of the potential is usually parameterized by the following slow-roll parameters,
\begin{align}
\varepsilon(\phi) &\equiv \frac{M_{pl}^2}{2} \left(\frac{V'}{V}\right)^2,\\
\eta({\f})  &\equiv M_{ pl}^2 \frac{V''}{V},
\end{align}
where the prime denotes the derivative with respect to $\phi$.
The slow-roll inflation takes place when $\varepsilon \ll 1$ and $|\eta| \ll 1$, and ends when one of the slow-roll
parameters becomes of order unity. Except for high-scale inflation, it is usually $\eta(\phi)$ that determines the end
of inflation.

The e-folding number between the horizon exit of the CMB pivot scale and the end of inflation
is given by
\begin{equation}
\label{eq:xecon}
N_* = \int^{\f_{\rm end}}_{\f_*}{\frac{H }{\dot{\f}} d \f} \simeq \int^{\f_{\rm end}}_{\f_*}{\frac{3H^2 }{-V'} d \f},
\end{equation}
where $H$ is the Hubble parameter, the overdot  represents the derivative with respect to time, and we have used the slow-roll
equation of motion, $3 H \dot{\phi} + V' \simeq 0$, in the second equality.  
Here the subscript ``$*$" represents the quantity evaluated at the horizon exit of the CMB scales.
Solving $|\eta| = 1$, we obtain the inflaton field value at the end of inflation,
\begin{align}
\label{phiend}
\phi_{\rm end} & \simeq 
\sqrt{\frac{2 \beta_n}{n^2-1}}
\frac{f^2}{M_{pl}} \ll \phi_{\rm min}.
\end{align}
Here $M_{pl} \simeq 2.4 \times 10^{18}$\,GeV is 
the reduced Planck mass. 
The e-folding number is evaluated as
\beq
\label{eq:einf}
N_*\simeq  \frac{V_0}{ 8  \lambda M^2_{pl}   \phi_*^2},
\eeq
where we have used $\phi_*\ll \phi_{\rm end}$.

The e-folding number is fixed once the inflation scale and the thermal history after inflation are given.  
As we will see in the next section the reheating takes place almost instantaneously after inflation
for $f \lesssim {\cal O}(10^8)$\,GeV, and in particular, this is the case for the inflaton within the reach of
ground-based experiments. 
By assuming the instantaneous reheating, the e-folding number is given by
\begin{equation}
\label{eq:efold}
N_*\simeq 28 -\log{\(\frac{k_* }{0.05\, {\rm Mpc^{-1}}}\)}+\log{\(\frac{V_0^{1/4} }{10\,\TEV}\)},
\end{equation}
where we have set the CMB pivot scale equal to $k_*=0.05\,{\rm Mpc}^{-1}$. Note that the e-folding number is about $30$ for
TeV-scale inflation, and it is much smaller than the conventionally adopted value, $N_*=50$ or $60$.

The inflaton acquires quantum fluctuations which perturb the evolution of the Universe depending on the position in space, inducing curvature perturbation. 
The predicted power spectrum of the curvature perturbation is given by
\begin{equation}
\label{eq:pln}
P_{\mathcal{R}}(k_*) 
\simeq 
\frac{V(\phi_*)^3}{12 \pi^2 V'(\phi_*)^2 M_{pl}^6},
\end{equation} 
while the observed the CMB normalization is
\begin{equation}
\label{pn}
P^{\rm CMB}_{\mathcal{R}}(k_*) \simeq 2.1 \times 10^{-9},
\end{equation}
at $k_* = 0.05\, {\rm Mpc}^{-1}$\cite{Akrami:2018odb}. 
Using Eqs.~(\ref{eq:lambda}) and \eq{einf}, we obtain
\beq
\label{eq:ratio}
\frac{\Lambda}{f}\simeq  1.2\times 10^{-3 }\(\frac{3}{n^2-1}\)^{\frac{1}{4}}\(\frac{30}{ N_*}\)^{\frac{3}{4}}.
\eeq
Thus, the ratio of the potential height and periodicity is fixed by the CMB normalization.

Interestingly, the CMB normalization alone fixes the relation between the inflaton mass and the decay constant
if $n$ is an even integer. In this case, the inflaton mass at the potential minimum $\phi = \phi_{\rm min}$
is of order the typical curvature scale, and given by
\begin{align}
m_\phi \simeq  \sqrt{2} {\frac{\Lambda^2}{f}}
\end{align}
when  $\theta\simeq 0$ and $\kappa \simeq 1$.
By using the relation \eq{ratio} we obtain~\cite{Czerny:2014xja}
\begin{align}
\label{mf}
m_\phi  \simeq  {2.1}\times 10^{-6} {\({\frac{3}{n^2-1}}\)^{1/2} \(\frac{30}{N_*}\)^{3/2}} f~~~[{\rm for~even}~n].
\end{align} 
Thus the inflaton mass scales in proportion to $f$. This relation is a rather robust prediction of the
axion hilltop inflation. As is clear from the above derivation, even if there are multiple cosine terms
that conspire to make the potential very flat around the maximum, a similar relation, $m_\phi \sim 10^{-6} f$, holds unless there are unnecessary cancellations among the parameters. This prediction
 can be tested by experiments if the inflation scale is sufficiently low.

If $n$ is an odd integer, on the other hand, the inflaton is massless in the limit of $\theta = 0$ and $\kappa = 1$,
and therefore, one needs to know the values of $\theta$ and $\kappa$ precisely to evaluate the inflaton mass. Their typical values can be inferred from the observed scalar spectral
index~\cite{Akrami:2018odb}
\beq
\label{nsCMB}
n_s^{\rm CMB}= 0.9649 \pm 0.0044,
\eeq
where we have adopted the result of $\it Planck$ TT,TE,EE+lowE.  
The scalar spectral index $n_s$ is given in terms of the slow-roll parameters as
\begin{equation}
\label{eq:nsform}
n_s \simeq  1- 6 \varepsilon + 2\eta.
\end{equation}
In the low-scale inflation, $\varepsilon(\phi_*)$ is much smaller than $|\eta(\phi_*)|$, and so, 
the spectral index is simplified to $n_s \simeq 1+ 2\eta(\phi_*)$. 
For $\theta = 0$ and $\kappa = 1$, 
we obtain 
\beq
n_s(\phi_*) \simeq 1-{\frac{3}{N_*}},
\eeq
which is too small to explain the observed value for $N_* \sim 30$. 
In fact, it is known that $n_s$ is rather sensitive to possible small corrections to the inflaton potential, and one can 
easily increase the predicted value of $n_s$ to give a better fit to the CMB data by introducing 
small but non-zero $\theta$~\cite{Takahashi:2013cxa, Czerny:2014wza, Daido:2017wwb, Daido:2017tbr}.
This is because a small CP phase $\theta$ contributes to the linear term of Eq.\,\eqref{app}, which slightly changes
the field value $\phi_*$ at the horizon exit in \Eq{xecon} for given $N_*$. 
Specifically, a small and positive $\theta$ reduces $\phi_*$ and $|\eta(\phi_*)|$, which increases the 
prediction of the spectral index. (Note that we assume $\dot{\phi}>0$ during inflation.) 
Introducing a nonzero $\kappa-1$ has a similar but slightly weaker effect~\cite{Takahashi:2013cxa}. 
In any case, one can realize $\eta(\phi_*) \sim -0.015$ 
to explain the observed $n_s$ by introducing a tiny $\theta (> 0)$ and/or $\kappa-1$. 
Then the inflaton acquires a nonzero mass at the minimum, and it is of order $\O(0.1-1) H_{\rm inf}$,
\begin{align}
   m_{\phi} = \O(10^{-7}-10^{-6}) 
\frac{1}{n} 
\frac{f^2}{M_{ pl}  } \(\frac{30}{N_*}\)^{3/2} ~~~[{\rm for~odd}~n].
\label{eq:mphi-f-oddn}
\end{align}
This is because the potential is upside-down symmetric as 
\beq 
\label{eq:udsym}V(\phi)=-V(\phi+\pi f )+{\rm const.,}
\eeq
which implies that the inflaton mass squared at the potential minimum is equal to the curvature at the potential maximum 
with an opposite sign.
Such light inflaton may be stable on the cosmological timescale and 
can be dark matter of the Universe. This is the so-called ALP miracle scenario~\cite{Daido:2017wwb, Daido:2017tbr}.\footnote{The unification of the inflaton and dark matter dates back to the seminal papers~\cite{Kofman:1994rk,Kofman:1997yn}.
See also Refs.~\cite{Mukaida:2014kpa,Bastero-Gil:2015lga,Chen:2017rpn,Hooper:2018buz,Borah:2018rca, Manso:2018cba,Rosa:2018iff,Almeida:2018oid,Choi:2019osi} for other recent studies.}
Interestingly, the viable parameter space will be probed by the IAXO experiment. Note that the inflaton mass at the potential minimum
is hardly changed if $n$ is an even integer, and the relation \eqref{mf} still remains valid even in the presence of small but nonzero $\theta$ and $\kappa-1$.

Before proceeding let us here make a rough estimate on typical values of $\theta$ and $\kappa - 1$ in the present 
axion hilltop inflation. The slow-roll inflation takes place if both slow-roll parameters are small, in other words,
if the first and second derivatives of the potential are extremely small. 
Since we have focused on the hilltop inflation 
rather than the inflection point inflation, 
the slow-roll inflation 
takes place around the potential maximum. The inflaton dynamics is considered to be dominated by the linear term
for a nonzero $\theta > 0$  introduced to explain the observed $n_s$. The inflaton field excursion during the 
last e-folding $N_*$ should not exceed $\phi_{\rm end}$. 
The limit on the inflaton field excursion and 
the slow-roll conditions around the potential maximum 
constrain the ranges of $\theta$ and $\kappa-1$
as
\beq
\label{range}
|\theta| \lesssim \(\frac{f}{M_{pl}}\)^{3}, |\kappa -1| \lesssim \(\frac{f}{M_{pl}}\)^{2}.
\eeq
For larger $|\theta|$ and $|\k-1|$, the slow-roll inflation no longer takes place around the potential maximum.
Instead, an inflection point inflation can happen where the inflation is driven by a cubic term that is obtained when expanded about the inflection point.\footnote{More precisely, the inflation ends due to the cubic term as we need to introduce
a linear term to explain the observed spectral index. 
See the discussion in Sec.~\ref{sec:5}.}
In this limit,
the location of the potential maximum moves toward negative values of $\phi$, and the inflaton dynamics will be
determined by the inflaton potential near the inflection point. This case will be discussed later.

  \begin{figure}[!t]
  \begin{center}
   \includegraphics[width=115mm]{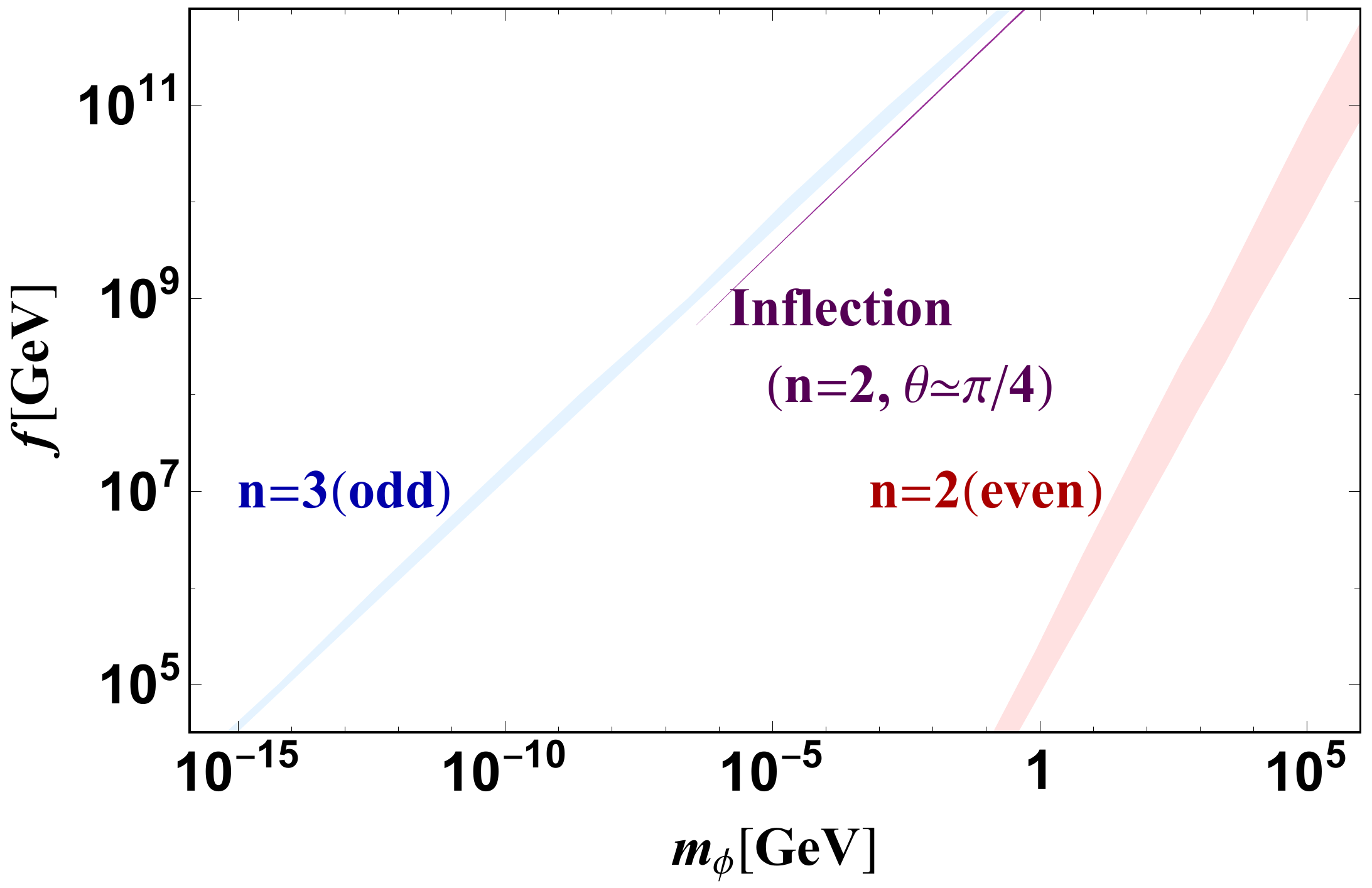}
\end{center}
\caption{
The relation between inflaton mass and the decay constant. The red band shows the even case with $n=2$. The blue band of $n=3$ and the purple line of inflection point inflation are also shown for comparison. }
\label{1}
\end{figure}

We have numerically solved the inflaton dynamics for different values of the decay constant $f$. 
For a given value of $f$, we vary $\theta$ and $\kappa$ in the ranges given in Eq.~(\ref{range})
to explain the observed CMB normalization (\ref{pn}) and the spectral index (\ref{nsCMB}). 
In addition, we have imposed the observed constraint on
the running of the spectral index~\cite{Akrami:2018odb},
\begin{align}
\label{eq:nsrun}
\frac{d n_s^{\rm CMB}}{d\log{k}}(k_*) & = -0.0045 \pm 0.0067, 
\end{align}
and on the running of running of the spectral index\cite{Cabass:2016ldu}, 
\begin{align}
\left|{\frac{d^2 n_s}{d\log{k}^2}}\right| <  0.001.
\end{align}
This is because the relatively small $N_*$ in the low-scale inflation 
may lead to a sizable scale dependence of the $n_s$.
In the numerical calculation, we keep the terms up to the third order of the slow-roll expansion
to evaluate the predicted spectral index and its running. 
We set the initial field value close enough to the potential maximum with a vanishing $\dot{\phi}$.\footnote{We have confirmed that our results do not change if we start from the potential maximum with $\dot{\phi}\sim H^2$, which is suggested from eternal inflation. (see Sec.~\ref{sec:5}.)}
The detailed analysis as well as the allowed ranges of $\k$ and $\theta$ can be found in Ref.~\cite{Daido:2017tbr}.

The relations  between $m_\phi$ and $f$ obtained in the numerical calculation
are shown in Fig.~\ref{1}. The red and blue bands correspond to the
cases of $n=2$ and $n=3$, respectively, where the bandwidth reflects the allowed ranges of 
$\theta$ and $\kappa$ as well as the observational uncertainties. 
 The two lines are our predictions based on the axion hilltop inflation.
 If the decay constant $f$ is small enough, it may be possible to probe the inflaton directly by various
ground-based experiments, which will be studied in the next section.

For comparison we also show the relation between $m_\phi$ and $f$ in the case of the inflection 
point inflation with $\theta \simeq \pi/4$ and $n=2$ (purple line), which is close to the case of $n=3$.
One can see that
 the decay constant is constrained to be greater than about $10^9\GEV$. This is because the running of the spectral index becomes too large for small $f$. In addition,
for the inflaton parameters within the reach of future experiments, the inflation scale turns out to be too small to reheat the Universe.

\section{Inflaton couplings to the SM and reheating}
\label{sec:3}
The slow-roll inflation ends when the potential becomes steep. After inflation ends, the Universe
will be dominated by the inflaton condensate oscillating around the vacuum $\phi_{\rm min}$.
To reheat the Universe the inflaton should transfer its energy to the SM particles through 
some interactions. As we shall see below, both perturbative decays and dissipation processes are
important to complete the reheating. 

The necessity of the inflaton couplings to the SM particles opens up the possibility of
producing inflaton at the ground-based experiments. This is indeed possible if the inflaton is 
kinematically accessible and
the couplings to the SM particles are strong enough. 
Then one may be able to check whether the inflaton mass and decay constant satisfy the relation \eqref{mf}.
In the following we will consider the two cases where the inflaton is coupled to photons (or weak gauge bosons)
or SM fermions. 
Through the two examples, we will study the prospect of the inflaton hunt at ground-based experiments and  the reheating process in detail.
Note that we will focus on the case with even $n$.
The detailed analysis of the case with odd $n$ can be found in Refs.~\cite{Daido:2017wwb,
Daido:2017tbr}.

\subsection{Inflaton coupling to photons}

Let us consider the inflaton/ALP coupling to photons, 
\begin{align}
\label{eq:int}
{\cal L} & = c_\g \frac{\a}{4 \pi} \frac{\phi}{f} F_{\mu \nu} \tilde F^{\mu \nu} 
\equiv \frac{1}{4} g_{\phi\g\g} \phi F_{\mu \nu} \tilde F^{\mu \nu},
\end{align}
 where
$c_\g$ is a model-dependent constant,
and $\alpha$ is the fine structure constant.
See e.g. Refs.~\cite{Jaeckel:2010ni,Ringwald:2012hr} for recent reviews on ALPs.
The ALP decays into a pair of photons through this interaction with the rate given by
\begin{equation}
\label{eq:dec}
\Gamma_{\rm dec,\g} =\frac{ c_\g^2\alpha^2 }{64 \pi^3} \frac{m_{\phi}^3}{f^2}.
\end{equation} 

\begin{figure}[!t]
  \begin{center}
   \includegraphics[width=120mm]{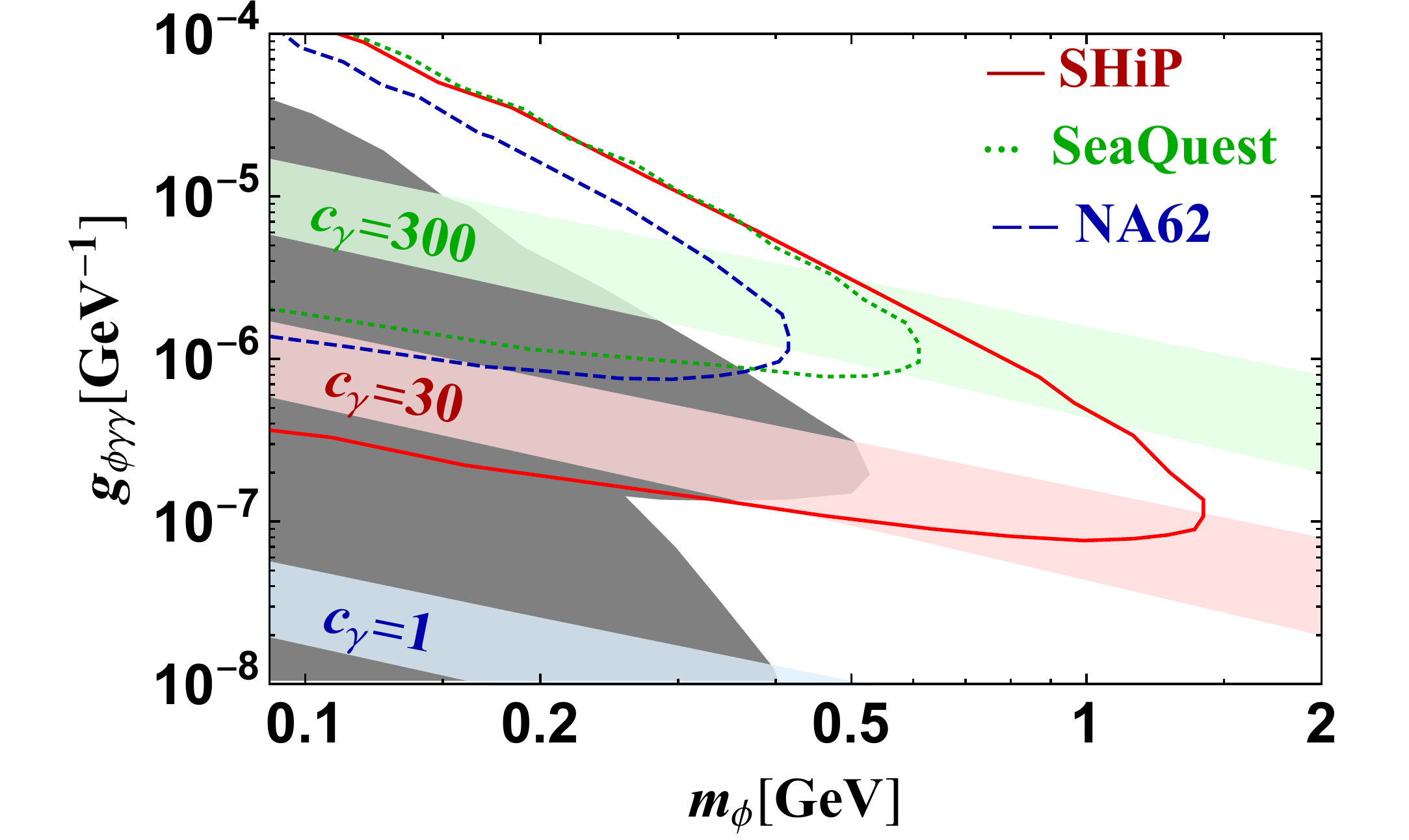}
\end{center}
\caption{
The parameter region predicted by the ALP inflation ($n=2$) with coupling to photons in the $m_{\f}-g_{\f\g\g}$ plane. The green, red, and blue bands correspond to 
$c_\g=300$, $30,\AND 1$, respectively, from top to bottom.  The red solid line denotes the projected sensitivity reach of the SHiP experiment.  The green dotted line and blue dashed line denotes the sensitivity of 
SeaQuest and NA62, respectively. The gray region is excluded by experiments and SN1987A. Both the sensitivity reach and excluded regions are adopted from Ref.~\cite{Dobrich:2019dxc}. }
\label{2}
\end{figure}

In Fig.~\ref{2} we show the parameter region predicted by the ALP inflation model with $n=2$ for $c_\g=1$ (blue band), $c_\g=30$ (red band) and $c_\g=300$ (green band). For comparison we also show the SHiP, SeaQuest, and NA62 (in beam dump mode) sensitivities in red solid, green dotted and blue dashed lines, respectively. 
The constraints from beam dump experiments and SN1987A are shown in gray region. The sensitivities as well as the constraints are adopted from Ref.~\cite{Dobrich:2019dxc}.\footnote{The possible reaches from other beam dump experiments by assuming certain environments can be found in Ref.~\cite{Harland-Lang:2019zur}. One can also find the sensitivity reaches of photon beam experiments in Ref.~\cite{Aloni:2019ruo}, where the authors also discuss the sensitivity with an ALP-gluon coupling. } 
One can see that the region predicted by the ALP inflation will be probed by the SHiP experiment if the anomaly coefficient is relatively large,
$c_\g \gtrsim 10$. For the $c_\g \gtrsim 100$, the region will be also tested by the experiments of SeaQuest~\cite{Berlin:2018pwi} as well as NA62~\cite{NA62:2017rwk}. The corresponding mass and decay constant are
$m_\phi = \O(0.1-1)\GEV$ and $f=\O(10^{4-6})\GEV$, respectively. In fact, the anomaly coefficient $c_\gamma$ can be easily of ${\cal O}(10)$. We shall see below a possible UV completion that leads to such value of $c_\gamma$.

Let us consider a UV completion a la the KSVZ axion model~\cite{Kim:1979if,
Shifman:1979if}. At temperatures above the electroweak scale  $T_{\rm EW}\simeq 100\GEV$, one should consider the ALP couplings to the weak gauge bosons as
\begin{align}
\label{eq:aWWBB}
{\cal L} & = c_2 \frac{\a_2}{8 \pi} \frac{\phi}{f}W^a_{\mu \nu} {\tilde W}_a^{\mu \nu}
+ c_Y \frac{\a_Y}{4 \pi} \frac{\phi}{f} B_{\mu \nu} \tilde B^{\mu \nu},
\end{align}
where $W^a_{\mu \nu}, \AND B_{\mu \nu} $ are the field strengths of SU(2)$_L$ and U(1)$_Y$, and $\alpha_2 = g_2^2/4\pi$ and $\alpha_Y=g_Y^2/4\pi$ are the  weak and hypercharge fine-structure constants, respectively.
Here $c_2$ and $c_Y$ are anomaly coefficients, and they are 
related to $c_\g$ as
\beq
c_\g= {\frac{c_2}{ 2}}+c_Y.
\eeq
Such anomalous couplings are induced if there are extra $N_\Psi$ fermion 
pairs $\Psi_i$ and $\bar{\Psi}_i$ in the representations
of $(Y_i, r_i, q_i), (-Y_i, r_i, q_i)$ under $({\rm U(1)}_Y, {\rm SU(2)}_L, {\rm U(1)}_{\rm PQ})$. Here U(1)$_{\rm PQ}$ is a global U(1) Peccei-Quinn
symmetry that is spontaneously broken, and  its associated Nambu-Goldstone boson (NGB) is the ALP (i.e., inflaton), $\phi$. In the simplest case, the ALP resides in the phase of a complex scalar field $S$ as
\begin{align}
    S = \frac{f}{\sqrt{2}} e^{i\phi/f},
\end{align}
where we have integrated out the heavy radial mode. One can couple $S$ to the fermion pair like $S \bar{\Psi}_i \Psi_i$ with $q_i = -1/2$ since
 the ALP transforms as (\ref{pq})
 under the U(1)$_{\rm PQ}$ transformation. It is possible that the ALP is given by a certain combination of the phases of multiple complex scalar fields. In this case the effective PQ charge $q_i$ can be greater than unity.
Through  one-loop diagrams with the fermions running in the loops, we obtain
\beq
c_2= \sum_i^{N_\Psi} 4T(r_i) q_i,~~ 
c_Y= \sum_i^{N_\Psi} 2 d(r_i)q_i Y_i^2.
\eeq
Here $T(r_i)$ is the index of the representation, e.g. $T(2)=1/2$ and $T(3)=2$, and $d(r_i)$ is the dimension of the representation, e.g. $d(2)=2$ and  $d(3)=3.$ 
For instance, one obtains $c_\g = 10 N_\Psi$ for  $q = 1$, $Y=1$, and  
$r=3$. So, $c_\gamma \geq 30$ is realized for $N_\Psi \geq 3$.
Note that some of the extra fermions are stable and their existence might lead to cosmological problems. In the following we assume that the extra fermions are so heavy that they are not produced after inflation.

Lastly let us note that, in principle,
 $c_Y$ and $c_2$ can be enhanced, while $c_\g$ remains small due to cancellation between the two. 
In this case, the ALP may be produced through the enhanced $\phi-\g-Z$ interaction, and it decays into photons 
at a displaced vertex~\cite{Bauer:2017ris, Bauer:2018uxu}. 
If the original relation between the mass and decay constant is retained in the photon coupling,
such a process may help us to probe the inflaton sector
and check the relation between the mass and decay constant.

\subsection{Reheating via photon coupling}

Now let us discuss the reheating through the photon coupling \eqref{eq:int}. We will find that the ALP within the
experimental reach reheats the Universe instantaneously after inflation via the dissipation effect. 

First, the ALP decays perturbatively into photons with the rate (\ref{eq:dec}), and the produced photons form thermal plasma.
The photons in thermal plasma acquire a thermal mass of order $eT$, and once it exceeds half the ALP mass, the perturbative decay 
becomes kinematically forbidden. This happens within one Hubble time after inflation if $ f \lesssim 10^9 c_\g^2 \GEV.$ 
 Then, the ALP condensate mainly evaporates through interactions with the ambient plasma such as $\f+\g \rightarrow e^{-}+e^{+}$~\cite{Yokoyama:2005dv,Anisimov:2008dz,Drewes:2010pf,Mukaida:2012qn,Drewes:2013iaa,Mukaida:2012bz}. 
Applying the result of Ref.~\cite{Moroi:2014mqa} to the ALP-photon coupling, 
we obtain the following dissipation rate, 
\begin{equation}
\label{eq:disIR}
\Gamma_{{\rm dis,\g}} = C  \frac{c_\g^ 2 \alpha^2T^3}{8 \pi^2f^2} \frac{m_\f^2}{e^4T^2},
\end{equation}
where $C$ is a numerical constant of ${\cal O}(1)$ which represents uncertainties of the
order-of-magnitude estimation of the dissipation rate and tachyonic preheating~\cite{Felder:2000hj,Felder:2001kt,Brax:2010ai}.\footnote{The tachyonic preheating makes the scalar condensate spatially inhomogeneous, enhancing the dissipation rate. The reheating will still be instantaneous for the parameters within the reach of the SHiP experiment.}
When the temperature becomes higher than the weak scale, 
the weak gauge bosons reach thermal equilibrium and the dissipation effect through the weak bosons becomes important. The rate is given by
\begin{equation}
\Gamma_{{\rm dis}} = C' \frac{c_2^2 \alpha_2^2T^3}{32\pi^2F^2} \frac{m_{\rm \f}^2}{g_2^4T^2}
+C'' \frac{c_Y^2 \alpha_Y^2T^3}{8 \pi^2F^2} \frac{m_{\rm \f}^2}{g_Y^4T^2}, \label{eq:disEW}
\end{equation}
where $C'$ and $C''$ are numerical constants of ${\cal O}(1)$.  

Now we evaluate the reheating temperature using the above decay and dissipation rates.
The ALP would evaporate completely if
\beq
\label{eq:evap}
\Gamma_{\rm dis} \gtrsim H.
\eeq
If this is satisfied soon after the inflation, the reheating is instantaneous and the reheating temperature is 
given by
\beq
T_R\simeq T_{\rm inst}\equiv \(\frac{30 }{\pi^2 g_*}V_0\)^{1/4}.
\eeq

Even if the dissipation rate initially is not high enough and \eq{evap} is not satisfied just after inflation, 
the dissipation becomes more efficient later
as $\Gamma_{\rm dis}$ decreases more slowly than $H$.
In this case, the reheating completes at $T=T_{\rm dis}$ when the dissipation rate becomes comparable to the Hubble parameter, $$ \Gamma_{\rm dis} \sim H=\sqrt{\frac{\pi^2 g_* }{90}}\frac{T_{\rm dis}^2}{M_{pl}}$$ 
and one obtains
\beq
\label{eq:upb}
T_{R}\simeq T_{\rm dis}\sim 100 \TEV \, C'' \(\frac{c_Y}{20}\)^2  \(\frac{30}{ N_*}\)^{3} \(\frac{ 3}{n^2-1}\).\eeq
Here we have assumed for simplicity that the second term of \eq{disEW} dominates over the first one.\footnote{Given $c_Y$, the reheating temperature in this case is determined by the CMB normalization.}

On the other hand, if the dissipation rate is sufficiently small, the plasma temperature may fall below the ALP mass, and the perturbative decay
becomes important again. This is the case if
$T_{\rm dis}$ defined above is smaller than the $m_{\phi}$.
The reheating completes when $H\sim \G_{{\rm dec},\g}$, and the reheating temperature is given by
\beq
T_R\simeq T_{\rm dec}\sim 100 \TEV   \( \frac{|c_\g|}{ 30}\)  \(\frac{ 3}{n^2-1}\)^{3/4}
\(\frac{30}{ N_*}\)^{9/4}
\(\frac{f}{10^{14}\GEV}\). 
\eeq
Here we have approximated that the main decay channel is a pair of photons for simplicity. 
To sum up, the reheating temperature is expressed as
\begin{align}
T_R\simeq \min{\Big[T_{\rm max} ,\max{\big[T_{\rm dis}, T_{\rm dec}\big]}\Big]}.
\end{align}

\begin{figure}[!t]
  \begin{center}
   \includegraphics[width=115mm]{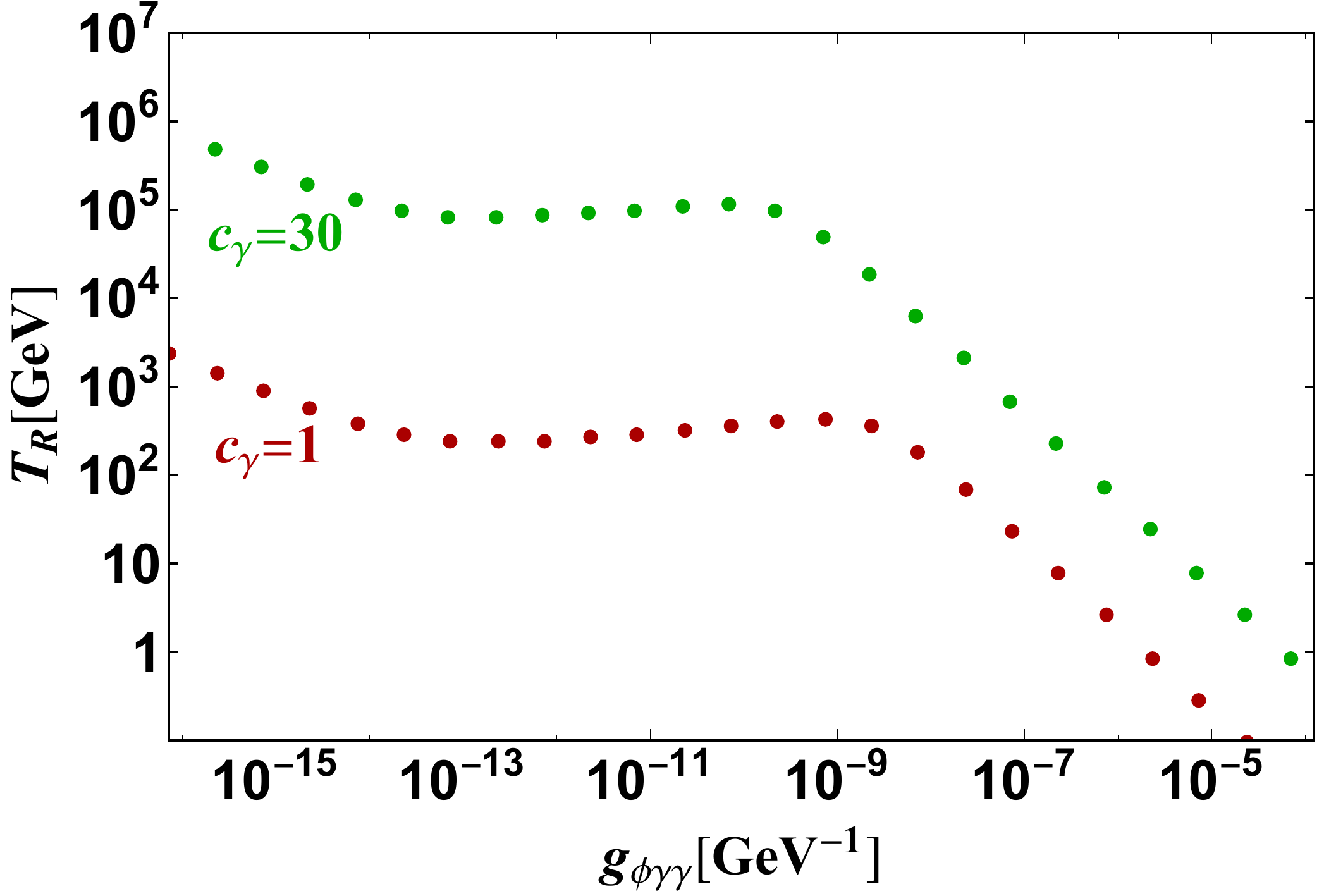}
\end{center}
\caption{
The reheating temperature 
of the ALP inflation ($n=2$) with coupling to photons is shown
as a function of $g_{\phi\g\g}$. 
Here we show the two cases with $c_\g=30$ ($c_Y=18$ and  $c_2=24$) [green] and with $c_\g=1$ ($c_Y=1$ and $c_2=0$)
[red].  
We take $C=C'=C''=1$ for simplicity.}
\label{4}
\end{figure}

To confirm the above na\"{i}ve estimate, we have numerically solved the Boltzmann equations,
\begin{align}
\left\{\begin{array}{ll}
	\displaystyle{\dot{\rho}_\phi+3H\rho_\phi=-\Gamma_{\rm tot}\rho_\phi} \\
	&\\
	\displaystyle{\dot{\rho}_r+4H\rho_r=\Gamma_{\rm tot}\rho_\phi}\label{evolution}
	\end{array}
	\right.,
\end{align}
where  $\rho_\phi$ and $\rho_r$ denote the energy density of the ALP condensate and that of radiation, respectively.  
The total decay/dissipation rate is given by 
\begin{align}
\G_{\rm tot}=\left\{
\begin{array}{ll}
\displaystyle{ \G_{\rm dec,\g}+\G_{\rm dis,\g}}~~~&{\rm for}~~ T<T_{\rm EW} \\ 
&\\
\displaystyle{ \G_{\rm dec,\g}+\G_{\rm dis,{\rm EW}}}~~~&{\rm for}~~ T>T_{\rm EW}
\end{array} 
\right.
\end{align}
If kinematically allowed, the perturbative decays to weak gauge bosons are also incorporated in our calculation.  The thermal blocking effect is included by using a step function. In the numerical calculation,
we evaluate the reheating temperature at
$\r_\phi=\r_\g$.
We show the reheating temperature in  Fig.~\ref{4}
as a function of $g_{\f\g\g}$.
The green (red) points represent the case for $c_\g=30~(1)$.
One can see from the figure that there is a plateau in the middle, which corresponds to the case where the reheating completes due to the dissipation process after inflation. As expected from Eq.~(\ref{eq:upb}), the reheating temperature does not depend on the decay constant. The reheating becomes instantaneous for larger $g_{\phi \g \g}$,
while the perturbative decay becomes important for smaller $g_{\phi \g \g}$. In particular, the reheating is instantaneous for the region within the sensitivity reach of SHiP, i.e., $g_{\phi\g\g}\simeq \O(10^{-7}-10^{-4})\GEV^{-1}$.
The reheating temperature is well above $\O(1)$ MeV
at which the big bang nucleosynthesis starts.

\subsection{Inflaton couplings to SM fermions}

Next we consider the inflaton (ALP) couplings to the SM fermions.\footnote{If one introduces additional fermions with certain couplings to the Higgs boson, the ALP may be produced at the LHC and future colliders~\cite{Bauer:2017ris, Bauer:2018uxu}. 
The relation of \eqref{mf} can be tested through the decay rates of the ALP into two photons or leptons. }
After the electroweak symmetry breaking, the couplings are given by
\beq
\laq{matbro}
{\cal L}=\sum_{i} i {\frac{c_{i} m_i}{f}} \f { \bar{\p}_i \g_5 \p_i},
\eeq
where $\p_i$ is a SM fermion with mass $m_i$, and $c_i$ is the sum of  the $\U(1)_{\rm PQ}$ charges of $\p_i$ and $\bar{\p}_i$.
In addition to the decay to the gauge bosons at the loop level, the ALP can decay into a pair of fermions at the tree-level if kinematically allowed.
The decay rate to the fermions is approximately given by
\begin{equation}
\label{eq:decf}
\G_{{\rm dec},\psi}\simeq
\sum_{m_{\p_i}<\frac{m_\phi}{2}}\frac{N_{\rm color}^{(i)} }{8\pi } \(\frac{c_i m_i }{f}\)^2m_{\rm \f},
\end{equation}
where $N_{\rm color}^{(i)}$ is the color factor of $\p_i$, and
the summation is over all the channels that are kinematically allowed.

\begin{figure}[!t]
  \begin{center}
   \includegraphics[width=120mm]{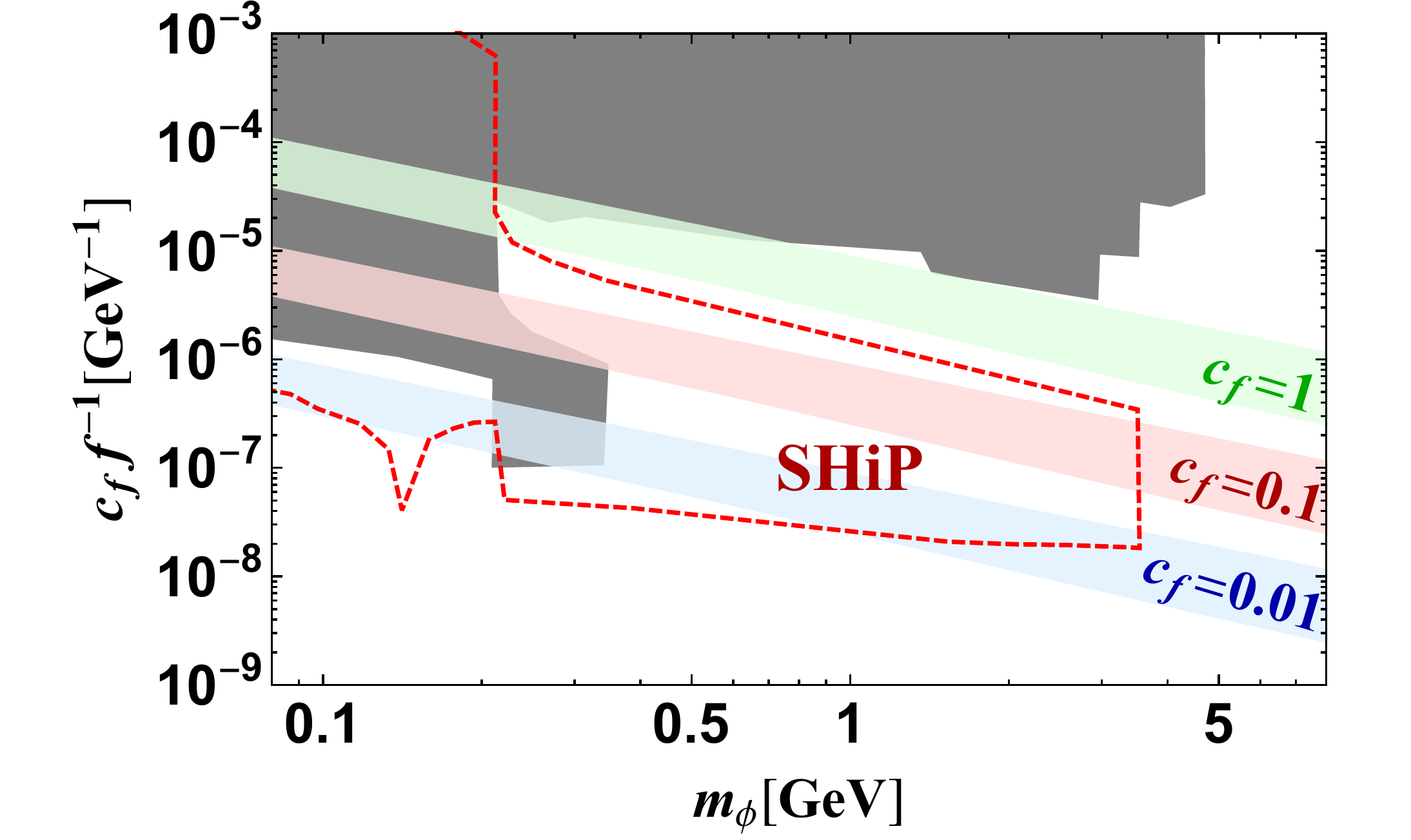}
\end{center}
\caption{
The parameter region predicted by the ALP inflation ($n=2$) 
with Yukawa-like couplings to all the SM fermions.
The green, red, and blue bands correspond to 
$c_f = 1$, 0.1, and 0.01, respectively from top to bottom.
The red dashed line denotes the sensitivity reach of the SHiP experiment, and the gray region is excluded by other experiments~\cite{Alekhin:2015byh}. }
\label{SHiP-fermion}
\end{figure}

ALPs with the couplings to the SM fermions 
can be searched for at the SHiP experiment~\cite{Anelli:2015pba,Alekhin:2015byh}.
In Fig.~\ref{SHiP-fermion} we show the parameter region 
predicted by the ALP inflation ($n=2$) with Yukawa-like couplings to all the SM fermions, together with the projected sensitivity reach of the SHiP experiment and the excluded regions, both of which are adopted from Ref.~\cite{Alekhin:2015byh}. Here we assume that the ALP has Yukawa-like couplings to all the SM fermions, i.e., $c_i$ takes a universal value $c_f$. One can see that a part of the predicted parameter region will be probed by the experiment if 
$c_f=\O(0.01-0.1)$. In fact, for $n\gtrsim 4$, the region with $c_f\gtrsim 1$ has an overlap with the sensitivity reach
because the ALP mass decreases as $n$ increases for given $f$. (See Eq.\,\eqref{mf}.)

Now let us consider a UV completion a la the DFSZ axion model~\cite{Dine:1981rt,
Zhitnitsky:1980tq}. We introduce two Higgs doublets, $H_u$ and  $H_d$,
which have PQ charges $Q_{H_u}$ and $Q_{H_d}$, respectively.
They develop vacuum expectation values (VEVs) at a temperature below $T_{\rm EW}$, $\vev{H_u^0}=v \sin(\b)$ and $\vev{H_d^0}=v \cos(\b)$, with the SM Higgs VEV $v\simeq 174\GEV$. 
The PQ charges satisfy $Q_{H_d}/Q_{H_u} = \tan^2 \beta$ to avoid a mixing of the ALP with the NGB eaten by the Z boson.
Integrating out the heavy Higgs bosons, the two Higgs doublets can be written in terms of the SM-like Higgs doublet, $H$, and the ALP as
\begin{align}
    H_u & = \tilde{H} \sin\b  \exp{\left[i Q_{H_u} \frac{\phi}{f}\right]},\\
    H_d & = H \cos\b  \exp{\left[i Q_{H_d} \frac{\phi}{f}\right]},
\end{align}
where $\tilde{H} = i \sigma_2 H^*$. By expanding the exponential with respect to the ALP, we obtain
\begin{equation}
    c_i = Q_{H_u} {\rm\, or~\,} Q_{H_d},
\end{equation}
depending on which Higgs the $\psi_i$ is coupled to. Note that the normalization of the PQ charge assignment is determined by
the transformation of the ALP given by (\ref{pq}). Therefore,
depending on the details of the PQ sector, especially how the ALP acquires the potential, $Q_{H_u}$ and $Q_{H_d}$ can easily take values of order $0.01$ to $100$ without invoking a very contrived set-up. In particular, if we couple only one of the Higgs doublets to the SM fermions, the strength of the couplings becomes universal, $|c_i| = c_f$, as assumed in Fig.~\ref{SHiP-fermion}.

\subsection{Reheating via matter coupling}

Now let us discuss the reheating via the couplings 
to the SM fermions. The ALP decays perturbatively 
into lighter fermions with the rate (\ref{eq:decf}). In addition, the ALP decays into photons and gluons
at the loop level. These perturbative decays populate
thermal plasma soon after the inflation ends. When the temperature of the plasma becomes as high as the ALP mass, 
 the back reaction becomes important and 
the perturbative decay is blocked. Afterward, the dissipation process 
such as $\phi+ \psi_i \rightarrow \psi_i + \gamma \,({\rm or~} g)$ becomes significant,  and its  rate is given by \cite{Mukaida:2012bz}
\begin{equation}
\label{eq:disbrF}
\G_{{\rm dis},\psi}\sim \sum_{m_i<T}\displaystyle{ N_{\rm color}^{(i)}\(\frac{c_i m_\psi}{f}\)^2 \(\frac{ \a_i}{2\pi^2}\) T,}
\end{equation} 
where we have omitted a numerical constant of $\O(1)$, and we set $\a_i=\a$ for leptons and $\a_i=\a_s$ for quarks. Here $\alpha_s$ is the strong coupling constant.

If the dissipation is efficient,  the temperature increases and when it becomes of order  $T_{\rm EW}$, the electroweak symmetry will be restored. 
At temperatures  higher than $T_{\rm EW}$, the Higgs boson is thermalized and it will participate in the dissipation process such as $\phi+h \rightarrow \p_i + \bar{\p}_i$ through the coupling, 
\beq
\laq{intsymfer}
\sum_{i} e^{i c_i \frac{\f }{f}}y_i \bar{\P}^L_{i} H \p^R_i + {\rm h.c.}\supset \sum_{i} i c_i y_i \frac{\f }{f} \bar{\P}^L_{i} H \p^R_i + {\rm h.c.},
\eeq
where $y_i= m_i/v.$ Here, $\bar{\P}_i^L$ and $ \p_i^R$ are, respectively, left-handed fermion doublet and right-handed singlet. For up-type Yukawa couplings, $H$ should be replaced with $\tilde{H}$.
We have neglected the flavor mixings for simplicity, and taken the leading term of $\f$ in the r.h.s.\footnote{The interactions in \Eq{intsymfer} can be cast into derivative interactions
plus anomaly couplings to gauge bosons after the chiral rotation of the fermion fields.  } This leads to the interaction of \eq{matbro} at the broken phase. 
The dissipation processes involving the Higgs field in the initial or final state are very efficient especially at high temperatures and its rate is given by\cite{Salvio:2013iaa} 
\begin{equation}
\label{eq:disF}
\G_{{\rm dis},\psi H}\sim
\sum_{i}N_{\rm color}^{(i)}  \(\frac{c_i^2 y_i^2  }{2 \pi^3 f^2}\) T^3.
\end{equation} 
Since the rate scales as $T^3$,
the dissipation that involves Higgs is most significant
when the temperature becomes higher than 
$T_{\rm EW}$ soon after the inflation.
The instantaneous reheating takes place if the following condition is satisfied,
 \beq\G_{{\rm dis},\psi H}\gtrsim H,\eeq
where the temperature is estimated by setting $V_0 = \pi^2 g_* T^4/30$ in the Hubble parameter (the right-hand side). This condition can be rewritten as
\beq
f\lesssim f_{\rm inst} = 2\times 10^{13}\GEV  c_t^{2} ({n^2-1})^{-1/4}\({\frac{40}{ N_*}}\)^{3/4}.
\eeq
Here we have assumed that the scattering process involving the
top quark, $i=t$, dominates the dissipation rate, and we have taken its $\overline{\rm MS}$ mass as $m_t\simeq 160\GEV$.

If $f\gtrsim f_{\rm inst}$, on the other hand, the dissipation involving the Higgs does not complete the reheating,
even though some fraction of the ALP condensate evaporates.
After reaching the maximum value, 
the dissipation rate starts to decrease faster than the Hubble parameter, and the remaining ALP condensate continues to dominate the Universe. This lasts until the ALP decays into
the SM fermions and the Higgs at the rate,
\beq
\G_{{\rm dec},\p H}\simeq \sum_{i}\frac{N_{\rm color}^{(i)} }{384\pi^3} c_i^2 y_i^2 \frac{m_\phi^3}{f^2},
\eeq
if kinematically allowed.
The reheating temperature corresponding to the 3-body perturbative decay is given by
\beq
T_R \simeq T_{\rm dec}=100\TEV |c_t| (n^2-1)^{-3/4} \(\frac{40}{N_*}\)^{9/4} \({\frac{ f}{ 10^{13}\GEV}}\)^{1/2},
\eeq
where we have assumed that the decay is dominated
by that into top quarks,
To sum up, one gets the reheating temperature expressed by
\beq
T_R\simeq 
\left\{
    \begin{array}{ll}
     \displaystyle{  T_{\rm inst}}~~~ 
     &{\rm for}~~f \lesssim f_{\rm inst}
     \\ 
    \displaystyle{ T_{\rm dec}}~~~ &
    {\rm for}~~f_{\rm inst}\lesssim f 
\end{array} 
\right..
\eeq
We note that, when $c_f$ is small, 
the 2-body decay can become important around $T=T_{\rm EW}$.
We do not discuss this case further since the relevant parameter region is not our focus.\footnote{When $c_f\ll 1$ and $ f\gtrsim f_{\rm inv}$, there is a parameter region where 
the ALP 2-body decay can prevent the temperature from falling below $T=T_{\rm EW}$. The temperature is kept at around $T_{\rm EW}$ for a while until the reheating completes and $T_R\sim T_{\rm EW}$. This behavior may lead to a non-trivial dynamics during the electroweak phase transition and may have implications for the electroweak baryogenesis. We leave a more detailed analysis for future work. }

  \begin{figure}[!t]
  \begin{center}
   \includegraphics[width=105mm]{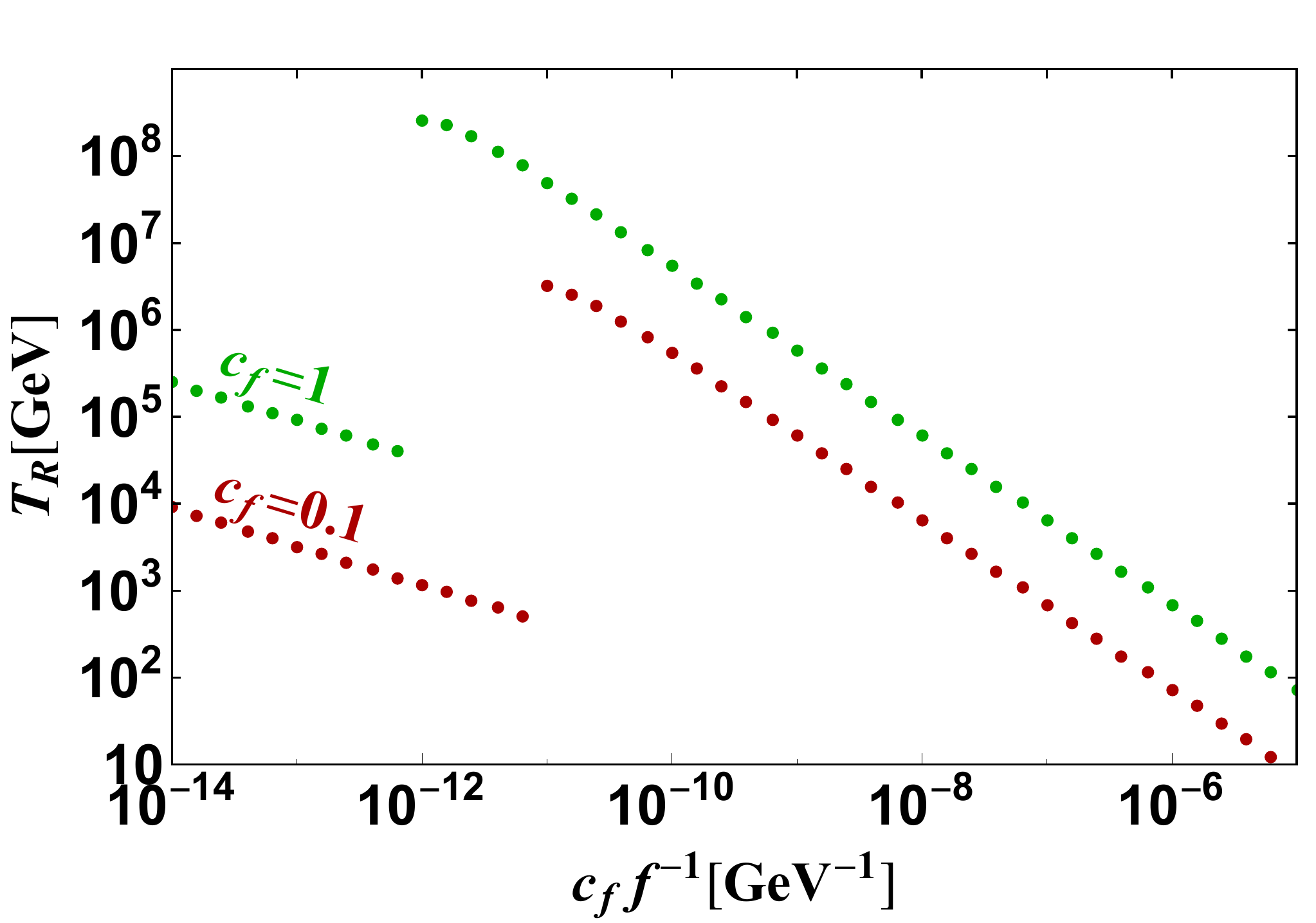}
\end{center}
\caption{
  The reheating temperature of the ALP inflation ($n= 2$)
  with coupling to the SM fermions as a function of, 
  $c_f^{-1} f$. The green (upper) and red (lower) points represent the cases with a universal coupling $c_f=1$, and $0.1$, respectively.  }
\label{5}
\end{figure}

We have numerically solved the Boltzmann equations \eqref{evolution} with
\begin{align}
\G_{\rm tot}=\left\{
\begin{array}{ll}
\displaystyle{ \G_{\rm dec}+\G_{\rm dis,\psi}}~~~&{\rm for}~~ T<T_{\rm EW} \\ 
&\\
\displaystyle{ \G_{\rm dec,\psi H}+\G_{\rm dis,\psi H}}~~~&{\rm for}~~ T>T_{\rm EW}
\end{array} 
\right.,
\end{align}
where we have assumed the universal couplings to the SM fermion, $c_i = c_f$, $\G_{\rm dec}$ is the sum of the 2-body and 3-body decay rates.
In Fig.~\ref{5}, the green (red) points represent $c_f=1 ~(0.1)$. In the region with $c_f f^{-1} > 10^{-12} ~(10^{-11}){\rm GeV}^{-1}$, the reheating is instantaneous. In particular,
this is the case in the ALP parameter
within the sensitivity reach of the SHiP experiment.

 \section{Inflection point inflation}
\label{sec:4}
So far we have focused on the hilltop inflation where the inflation takes place near the potential maximum and the quartic coupling terminates the inflation.
In fact, there is another possibility that 
inflation takes place at an inflection point that is not necessarily close to the potential maximum.
In particular, it is the cubic term that terminates the inflation. The purpose of this section is to study the dynamics of the inflection point inflation in our set-up, and to see that it is qualitatively different from the hilltop inflation.

The inflection point inflation takes place in the vicinity of
a flat inflection point $\phi = \phi_{\rm IP}$ 
where $V'=0$ and  $V''= 0$ are satisfied. These two conditions can be satisfied at any $\phi_{\rm IP}$ for a certain choice of the two parameters $\theta$ and $\kappa$. Except for $\phi_{\rm IP}$ very close to the potential maximum, $\theta$ is generically of order unity. The third and fourth derivatives of the potential at the inflection point are considered to be of order $\Lambda^4/f^3$ and $\Lambda^4/f^4$, respectively. By expanding the potential around $\phi = \phi_{\rm IP}$, one obtains
\beq
\label{eq:infl}
    V\simeq V_0 - c_1 \frac{\L^4}{f^3} \hat{\phi} ^3 - c_2 
\frac{\L^4}{ f^4}\hat{\phi}^4 ,
\eeq
where $\hat{\phi}\equiv \f-\phi_{\rm IP}$, and
$c_1$ and $c_2$ are functions of $\h$ and $\k$. 
For instance, there is an inflection point at
$\phi_{\rm IP}=\pi f/4$ for $n=2$, $\theta = \pi/4$,
and $\kappa = 2$, leading to $c_1=1/2,\AND c_2=0$. The inflaton potential in this case is shown in the right panel of Fig.~\ref{pot}.
Unless $c_1$ is extremely suppressed as in the case of the
hilltop inflation, the cubic term dominates over the quartic one. By neglecting the quartic term, one can solve \Eq{xecon}
to obtain the inflaton field value at the horizon exit
\beq
\hat{\phi}_*\simeq \frac{ V_0 f^3}{3 c_1 \L^4  M_{pl}^2 N_*}.
\eeq
The spectral index is given by
\beq 
n_s\simeq 1-\frac{4}{N_*},
\eeq
which is smaller than the observed $n_s$ (\ref{nsCMB}).

One can increase the predicted spectral index by introducing a small linear term, $$\d V = -\hat{\theta}  \frac{\L^4}{ f} \hat{\phi}$$  to (\ref{eq:infl}). Here $\hat{\theta}$ is a function of $\theta \AND \k$, and is obtained by slightly changing the value of $\theta\AND \k$ from their canonical values to obtain the (flat) inflection point. A positive $\hat{\theta}$ reduces $\phi_*$, thereby increasing the predicted value of $n_s$.
Note here that the linear potential itself does not contribute to the $\eta$ parameter.
To increase the spectral index by a sufficient amount, the linear term should give a dominant contribution to $V'$ 
when the CMB scales exited the horizon. In other words,
\beq 
\hat{\theta}\sim c_1^{-1} N_*^{-2} \(\frac{f}{M_{pl}}\)^4>0. 
\eeq
Here and hereafter, we approximate $V_0\sim \L^4.$ As we will see in the next section,  eternal inflation does not take place around the inflection point due to the linear term.

Now let us estimate the inflaton mass at the minimum and the inflation scale.
From the CMB normalization, one gets
\beq
m_\phi \sim \frac{\L^2}{f}\simeq {0.1}\EV  
\, c_1^{-1} N_*^{-2}
\(\frac{f}{10^{6}\GEV} \)^2.
\eeq
For $c_1=\O(1)$, the inflaton mass is much smaller than that 
for hilltop inflation with even $n$.
In fact, this relation between $m_\phi$ and $f$ is very similar to (\ref{eq:mphi-f-oddn}) for the hilltop inflation with odd $n$. This explains why the line of the hilltop inflation with $n=3$ is close to that of the inflection point inflation in Fig.~\ref{1}. In fact,
despite the similarity in the relation between the inflaton mass and decay constant, the inflation scale is very different between the two cases. 
This can be understood by noting that
the light inflaton mass is realized by
cancellation between the two cosine terms in the case of the hilltop inflation with odd $n$. Such a cancellation was a direct outcome of requiring the slow-roll inflation around the potential maximum. On the other hand, the inflaton mass is given by its typical value on dimensional grounds in the case of the inflection point inflation, and so,
 the inflation scale $(V_0)^{1/4}\sim \L$ is suppressed as
\beq
V_0^{1/4}\sim ({ m_\phi f})^{1/2}\sim 10\MEV \, c_1^{-1/2} {N_*^{-1}} \(\frac{f}{10^6\GEV}\)^{3/2}.
\eeq
As a result, successful reheating is only possible for relatively large values of $f$,
well beyond the experimental reach. For instance, by coupling the ALP to photons one gets  
the reheating temperature from the perturbative decay as
\beq
 T_{\rm dec}\sim  10\MEV c_\g c_1^{-3/2} N_*^{-3} \(\frac{f}{10^{13}\GEV}\)^2.
\eeq
One may enhance the coupling to photons by the clockwork mechanism~\cite{Higaki:2015jag,Farina:2016tgd}, 
but the original $m_\phi-f$ relation will then be difficult to check.

The small inflation scale also tends to generate too large running of the spectral 
index, ${d n_s/ d \log k}\simeq -4/N_*^2 $. (See \Eqs{efold} and \eq{nsrun}, 
and e.g. Ref.~\cite{Daido:2017tbr} for the derivation.)
The viable parameter region of the inflection point inflation with 
$\theta=\pi/4+\d\theta, \k=2+\d\k, n=2 $ is shown in Fig.~\ref{1}. 
We vary $-0.3<\d\theta ({f/ M_{pl}})^{-2} <0.3$ and  $-0.1<\d \kappa ({f/M_{pl}})^{-4}<0.1$ to explain the CMB data.
We find that for $f\lesssim 10^9\GEV$ the prediction is not consistent with the CMB data due to the 
too large running of the spectral index.
We note however that the running of the spectral index may be alleviated by introducing a small 
modulation to the potential~\cite{Kobayashi:2010pz}. 

Lastly let us comment on the choice of the initial condition, which is somewhat related to the topic of
the next section. In fact, it is not clear what the initial condition of the inflaton field is appropriate 
since the potential does not allow eternal inflation as we shall see shortly.
In the numerical simulation, we set the vanishing initial velocity $\dot{\phi}=0$, and start from
the inflaton field close enough to the inflection point so that the slow-roll inflation takes place. 
If the initial field value is away from the inflection point the inflaton would overshoot the 
inflection point without slow-roll. For a certain choice of the initial condition, the inflaton
experiences a short period of ultra slow-roll~\cite{Tsamis:2003px,Kinney:2005vj, Dimopoulos:2017ged}
before entering the slow-roll regime. However, for most of the parameter space, 
the CMB scales exit the horizon during the slow-roll regime, and the presence of the ultra slow-roll
does not affect the observation.\footnote{
We thank K. Dimopoulos for a useful comment on the ultra slow-roll.  }

\section{Eternal inflation}
\label{sec:5}
A most important advantage of the axion hilltop inflation is that it incorporates eternal inflation~\cite{steinhardt-nuffield,
Vilenkin:1983xq,
Linde:1986fc,
Linde:1986fd,
Goncharov:1987ir,
Guth:2007ng}. If the ALP initially stays close enough to the potential maximum, its quantum fluctuation dominates over the classical motion. Both of the slow-roll parameters are smaller than unity as long as $\theta$ and $\kappa$ satisfy (\ref{range}), and therefore, eternal inflation takes place. Once the eternal inflation starts, it may compensate for possible fine-tuning of the parameters to realize the low-scale inflation. Moreover, low-scale eternal inflation has interesting implications for the QCD axion abundance~\cite{Graham:2018jyp, Guth:2018hsa} as well as the moduli problem~\cite{Ho:2019ayl}

On the other hand, it is difficult to realize the eternal inflation in the case of the inflection point inflation
at least in its minimal form. To see this, let us first note that the linear term with $\hat{\h}>0$ dominates $V'$ around the inflection point. Then, the classical motion of the inflaton over the Hubble time
 is given by,
 \beq 
\Delta \phi_{\rm classical} = H^{-1} \dot{\f}_{\rm cl} \sim 
H^{-2}\frac{\L^4}{3 f} \hat{\theta},
\eeq
while the quantum fluctuation is given by
\beq
\Delta \phi_{\rm quantum} = \frac{H }{2\pi}.
\eeq
Thus, one finds that 
\beq \frac{\Delta \phi_{\rm quantum}}{\Delta \phi_{\rm classical}}\sim 10^{-5}, 
\eeq
namely, the classical motion always dominates over the quantum diffusion even at the inflection point. 
Therefore, eternal inflation does not take place due to the linear term added to increase $n_s$. Note that, since $|\theta|$ and $|\kappa-1|$ become larger than the upper bound given by (\ref{range}), 
the slow-roll condition is not satisfied at the potential maximum, in contrast to the hilltop inflation. Even if one sets the initial inflaton field value near the potential maximum,
it would soon start to roll down and overshoot the inflection point.
Therefore, eternal inflation is not realized in the case of the inflection point inflation. In this sense, the amount of the fine-tuning of the parameters to realize low-scale inflation may have to be taken at a face value in this case. In order to implement eternal inflation before the inflection point inflation kicks in, we need to change the potential slightly to have a local minimum somewhere, or introduce another sector to drive eternal inflation.

\section{Discussion and conclusions}
\lac{5}

We have so far mainly focused on the potential with two cosine terms like Eq.~(\ref{eq:DIV}), but the predicted relation between the inflaton mass and decay constant \eqref{mf} holds in a more general case. This is because the relation \eqref{mf} is derived only under the assumption that the inflation takes place around the potential maximum approximated by
\beq 
V \;\simeq\; \Lambda^4-\l\f^4 + \cdots
\eeq
with 
\beq
\l \sim \frac{\L^4}{f^4},
\eeq
which is fixed by the CMB normalization to be about $10^{-12}$, and that the inflaton mass at the minimum is given by
\beq
m^2_\f \sim \frac{\L^4}{f^2}
\eeq
on dimensional grounds.
The condition for the slow-roll inflation is essentially to suppress the quadratic term in the potential. Therefore, unless extra fine-tuning is imposed, our prediction \eqref{mf} generically holds without significant changes even for the inflaton potential that consists of multiple cosine terms with different $n$s.
One exception is the case with the inflaton potential that consists of only cosine terms with odd $n$s.
In this case, the inflaton is much lighter than the naively estimated value as in the ALP miracle scenario\cite{Daido:2017wwb,
Daido:2017tbr}.

One may wonder whether the flat-top inflaton potential is stable against radiative corrections induced by the couplings introduced for reheating. To be concrete, let us consider the ALP-photon interaction \eq{int}. 
This interaction preserves a continuous shift symmetry of ALP and the radiative correction from \eq{int} itself does not contribute to the non-zero modes of the cosine terms in the ALP potential. 
However, the introduction of the \eq{int} may alter the instanton effect generating the potential \eq{DIV}. The correction is of order 
$$ \d V\sim \frac{g_{\f\g\g}^2}{(16\pi^2)^2} \frac{\L^8  }{f^4}  \cos{\(\frac{n \phi}{f}\)}$$ which arises from the two loop diagram including an instanton and two vertices of \eq{int}. 
This correction changes the relative height $\kappa$ by about $g_{\phi\g\g}^2 \L^4/ (16\pi^2f)^2$. Such correction is 
negligible if it is smaller than $(f/M_{pl})^2$. This is the case if $f/c_\gamma \gtrsim 10^7\GEV$. Even for $f/c_\gamma< 10^7\GEV$, one can redefine $\k-1$ to absorb the correction and then our discussion remains intact. Importantly, once the relative height is given, it does not run by changing the renormalization scale, thanks to the periodicity guaranteed by the unbroken discrete shift symmetry.

Let us discuss implications of the ALP inflation for dark matter. For the parameters that can be 
probed by the ground-based experiments, the  
reheating is instantaneous and the reheating temperature
is as high as $\Lambda \sim 10^{-3} f ={\cal O}(10 - 10^4)$\,GeV. Then, a WIMP with mass up to $\O(10) \Lambda$  reaches thermal equilibrium and can explain the observed dark matter abundance for certain annihilation processes. 
Therefore, WIMP is a good dark matter candidate. 

 On the other hand, the QCD axion~\cite{Graham:2018jyp, Guth:2018hsa} and string axions~\cite{Ho:2019ayl} will follow the so-called Bunch-Davies distribution peaked at the potential minimum during the low-scale eternal inflation.
Then, with the Hubble parameter during inflation
less than keV, the abundance of the QCD axion and string axions will be too small to explain  dark matter. Here the assumption is that the minimum of the axion potential 
remains unchanged (modulo the potential period)  during and after inflation~\cite{Guth:2018hsa,Ho:2019ayl}. 
In fact, if the inflaton has a mixing with the axion, the  minimum of the axion potential can be shifted after inflation, inducing coherent oscillations of the axion at a later time when the Hubble parameter becomes comparble to the axion mass. To see this let us consider
a case with the Yukawa-like couplings $c_i=c_f$, where
the inflaton coupling to gluons is radiatively induced,
\begin{align}
    {\cal L} = \frac{\alpha_s}{8 \pi}\left(
    6 c_f \frac{\phi-\phi_{\rm min}}{f}+\frac{a}{f_a}
    \right)G_{\mu\nu}^b\tilde{G}_b^{ \mu \nu},
\end{align}
where $a$ denotes the QCD axion, $f_a$ is its decay constant, $G_b^{\mu \nu}$ is the field strength of $\SU(3)_c$, and we have chosen a constant CP phase so that the QCD axion
has a low-energy minimum at $a=0$. If the Hubble parameter during inflation
is smaller than the QCD scale, the above combination of $\phi$ and $a$ acquires a mass from non-perturbative QCD effects. For simplicity we assume that this mass scale is much smaller than the curvature of inflaton potential so that $a$ approximately coincides with the lighter mass eigenstate.\footnote{
If not, the inflaton will be given by a certain combination of $\phi$ and $a$, and the inflaton dynamics will be similar to a hybrid inflation considered in Ref.~\cite{Daido:2017wwb}.
}  The inflaton $\phi$ stays near the origin, $\phi \approx 0$, during inflation, while it is stabilized at $\phi = \phi_{\rm min} \approx \pi f_\phi$ after inflation. If the inflation lasted sufficiently long, then, the probability distribution of the QCD axion follows the Bunch-Davies distribution peaked at $a/f_a \approx 6 \pi c_f$ (mod $2\pi$). This sets the typical initial oscillation amplitude of the QCD axion. For instance,
if $6\pi c_f \sim 1$, the QCD axion with $f_a = {\cal O}(10^{12})$\,GeV can account for dark matter.

Lastly let us mention a possible connection between the ALP decay constant $f_\phi$ and the gravitino mass $m_{3/2}$. This is based on the observation that soft SUSY breaking effects can induce
spontaneous breaking of U(1)$_{\rm PQ}$ symmetry.
To see this we introduce two PQ chiral superfields, $S$ and $\bar{S}$,
with the following superpotential,
\begin{align}
    &W\supset \frac{\lambda_S}{2}  S^2 \bar{S}
\end{align}
where  $\lambda_S$ is a coupling constant of order unity.
For a general K\"ahler potential, the lowest components of $S$ and $\bar{S}$ acquire soft SUSY breaking masses of order $m_{3/2}$ through
supergravity effects and Planck-suppressed couplings to the SUSY breaking field $Z$ with $F_Z \simeq m_{3/2} M_{ pl}$. If the soft mass of $S$ ($\bar{S}$) is negative (positive), $S$ develops a nonzero VEV while $\bar{S}$ stays near the origin,
\begin{align}
   \vev{S}\simeq \frac{ m_{3/2}}{\lambda_S},~ \vev{\bar{S}}\simeq 0. 
\end{align}
Therefore, if the phase of $S$ is identified with the ALP, the decay constant $f_\phi$ is naturally of order $m_{3/2}$.
Interestingly, $m_{3/2}\simeq 10-1000\TEV$ is favored from the SM Higgs boson mass.  For instance, in the pure-gravity mediation \cite{Ibe:2006de,Ibe:2011aa,Ibe:2012hu} or minimal split SUSY~\cite{ArkaniHamed:2012gw} (see also Refs.~\cite{Wells:2003tf,
Arvanitaki:2012ps, Hall:2012zp}), the stop masses of order $m_{3/2}
=\O(10-100)\TEV$
is favored   to explain the Higgs boson mass. In the Higgs or Higgs-anomaly mediation~\cite{Yamaguchi:2016oqz, 
Yin:2016shg,Yanagida:2016kag,Yanagida:2018eho
}, the stop mass squared is generated at the loop level and can be of order $0.01m_{3/2}^2$. The favored range of the gravitino mass is $m_{3/2}=\O(100-1000)\TEV.$
We emphasize here that, when combined with the relation between the ALP mass and decay constant (\ref{mf}),  the relation $f\sim m_{3/2} = {\cal O}(10^{4-6})\GEV$ implies that  the sensitivity reach of the SHiP experiment covers a large fraction of the parameter space of the ALP inflation. 
Also, if the gaugino masses are generated at the loop level, e.g. anomaly mediation~\cite{Giudice:1998xp, Randall:1998uk},\footnote{
Strictly speaking, the gaugino masses may be generated from deflected anomaly mediation~\cite{Pomarol:1999ie,Rattazzi:1999qg} due to the contributions of the PQ fields.
} 
\beq
M_i \sim (10^{-2}-10^{-3}) m_{3/2}
\eeq
thermal relic of the lightest neutralino can explain dark matter. Notice that this mass range is below $\O(10)T_R\sim 10^{-2}m_{3/2}$ so that the neutralino will reach thermal equilibrium before freeze-out. 
For instance, the wino with $M_2\simeq 
2.7-3\TEV$ can explain the observed dark matter abundance. 
Note also that the cosmological moduli and gravitino problems, 
the two notorious cosmological problems in SUSY, 
are greatly relaxed in this scenario thanks to very low inflation scale and reheating temperature. 
If the inflaton is discovered in beam dump experiments, 
its decay constant point to an energy scale at which new physics appears. It may be high-scale SUSY which is consistent with the observed
Higgs boson mass.

In this paper we have discussed the ALP inflation model where 
the inflaton potential consists of two (or more) cosine terms,
and showed that there is a robust relation between the inflaton mass at the potential minimum and the decay constant, $m_\phi \sim 10^{-6} f$, for a class of the axion hilltop inflation models. For successful reheating, the inflaton (or ALP) must have sizable couplings to the SM particles, especially if the inflation scale is low. Such large couplings can be introduced  without spoiling the flatness of the potential by radiative corrections thanks to the discrete shift symmetry. 
We have studied the reheating process through both perturbative decays and dissipation effects for the two cases of the ALP couplings to photons (or weak gauge bosons) and the SM fermions. In both cases, the reheating is instantaneous for the parameters within the experimental reach. If the ALP with the predicted relation between the mass and decay constant is found by experiments such as the SHiP experiment, it is possible that the ALP played the role of the inflaton at the very early stage of the history of the Universe.
If so, the decay of such particle observed in the experiment proceeds through the same interaction that reheated the Universe. In some sense, the Big Bang may be probed at experiments on earth.

\section*{Acknowledgments}
FT thanks H. Matsui for discussion on eternal inflation. WY would like to thank the Particle Theory and Cosmology Group at Tohoku University for warm hospitality during the completion of this work.
This work is supported by JSPS KAKENHI Grant Numbers
JP15H05889 (F.T.), JP15K21733 (F.T.), 
JP26247042 (F.T),  JP17H02875 (F.T.), 
JP17H02878(F.T.), by NRF Strategic Research Program NRF-2017R1E1A1A01072736 (W.Y.), and by World Premier International Research Center Initiative (WPI Initiative), MEXT, Japan.

\end{document}